\documentclass[english,review]{elsarticle}
\usepackage{mathpazo}

\usepackage[T1]{fontenc}
\usepackage[latin9]{inputenc}
\usepackage{geometry}
\usepackage{color}
\usepackage{array}
\usepackage{units}
\usepackage{multirow}
\usepackage{graphicx}
\usepackage{esint}

\makeatletter

\providecommand{\tabularnewline}{\\}

\makeatother

\usepackage{babel}
\begin{document}

\title{Fluid preconditioning for Newton-Krylov-based, fully implicit, electrostatic
particle-in-cell simulations }

\author[lanl]{G. Chen\corref{cor1}}

\ead{gchen@lanl.gov}

\author[lanl]{L. Chacón}

\author[boulder]{C. Leibs}

\author[lanl]{D. Knoll}

\author[UNM]{W. Taitano}

\cortext[cor1]{Corresponding author}

\address[lanl]{Los Alamos National Laboratory, Los Alamos, NM 87545}

\address[boulder]{University of Colorado Boulder, Boulder, CO 80309}

\address[UNM]{University of New Mexico, Albuquerque, NM 87131}
\begin{abstract}
A recent proof-of-principle study proposes an energy- and charge-conserving,
nonlinearly implicit electrostatic particle-in-cell (PIC) algorithm
in one dimension {[}Chen et al, \emph{J. Comput. Phys.}, \textbf{230}
(2011) 7018{]}. The algorithm in the reference employs an unpreconditioned
Jacobian-free Newton-Krylov method, which ensures nonlinear convergence
at every timestep (resolving the dynamical timescale of interest).
Kinetic enslavement, which is one key component of the algorithm,
not only enables fully implicit PIC a practical approach, but also
allows preconditioning the kinetic solver with a fluid approximation.
This study proposes such a preconditioner, in which the linearized
moment equations are closed with moments computed from particles.
Effective acceleration of the linear GMRES solve is demonstrated,
on both uniform and non-uniform meshes. The algorithm performance
is largely insensitive to the electron-ion mass ratio. Numerical experiments
are performed on a 1D multi-scale ion acoustic wave test problem. \end{abstract}
\maketitle

\section{Introduction}

The Particle-in-cell (PIC) method solves Vlasov-Maxwell's equations
for kinetic plasma simulations \citep{birdsall-langdon,hockneyeastwood}.
In the standard approach, Maxwell's equations (or in the electrostatic
limit, Poisson equation) are solved on a grid, and the Vlasov equation
is solved by method of characteristics using a large number of particles,
from which the evolution of the probability distribution function
(PDF) is obtained. The field-PDF description is tightly coupled. Maxwell\textquoteright{}s
equations (or a subset thereof) are driven by moments of the PDF such
as charge density and/or current density. The PDF, on the other hand,
follows a hyperbolic equation in phase space, whose characteristics
are self-consistently determined by the fields.

To date, most PIC methods employ explicit time-stepping (e.g. leapfrog
scheme), which can be very inefficient for long-time, large spatial
scale simulations. The algorithmic inefficiency of standard explicit
PIC is rooted in the presence of numerical stability constraints,
which force both a minimum\textcolor{black}{{} grid-size (due to the
so-called finite-grid instability \citep{birdsall-langdon,hockneyeastwood},
which requires resolu}tion of the smallest Debye length)\textcolor{black}{{}
and a very small timestep (due to the well-known CFL constraint, which
requires resolution of the fastest plasma wave, or in the more general
electromagnetic case, the light wave)}. Moreover, a fundamental issue
with explicit schemes is numerical heating due to the lack of exact
energy conservation in a discrete setting \textcolor{black}{\citep{birdsall-langdon,hockneyeastwood},
which makes the accuracy of explicit PIC simulations questionable
on long time scales. }This problem is particularly evident for realistic
ion-to-electron mass ratios.

Implicit methods hold the promise of \textcolor{black}{overcoming
the difficulties and inefficiencies of explicit methods for long-term,
system-scale simulations. Exploration of implicit PIC started in the
1980s. Two approaches, namely implicit moment \citep{mason-jcp-81-im_pic,brackbill-forslund}
and direct implicit \citep{friedman-cppcf-81-di_pic,friedman1991multi}
methods, were explored. Both approaches use linear implicit schemes
to simplify the inversion of the original Vlasov-Poisson-Maxwell system,
and both enable a numerically stable time integration with large timesteps.
These implicit approaches avoided inverting the system of a large
set of coupled field-particle equations by using a predictor-corrector
strategy. The main limitation of these linear implicit schemes is
the lack of }nonlinear convergence, which leads to inconsistencies
between fields and particle moments. As a result, significant numerical
heating is often observed in long term simulations \citep{cohen-jcp-89-di_pic}.

There has been significant recent work exploring fully implicit, fully
nonlinear PIC algorithms, either Picard-based \citep{taitano-sisc-13-ipic}
(following the implicit moment method school) or using Jacobian-Free
Newton-Krylov (JFNK) methods \citep{chen-jcp-11-ipic,markidis2011energy}
(more aligned with the direct implicit school). In contrast to earlier
studies, these nonlinear approaches enforce nonlinear convergence
to a specified tolerance at every timestep. Their fully implicit character
enables one to build in exact discrete conservation properties, such
as energy and charge conservation \citep{chen-jcp-11-ipic,taitano-sisc-13-ipic}.
In these studies, particle orbit integration is sub-stepped for accuracy,
and to ensure automatic charge conservation.

The purpose of this study is to demonstrate the effectiveness of fluid
(moment) equations to accelerate a JFNK-based kinetic solver (moment
acceleration in a Picard sense has already been demonstrated in Ref.
\citep{taitano-sisc-13-ipic}). An enabling algorithmic component
of the JFNK-based algorithm is the enslavement of particles to the
fields, which removes particle quantities from the dependent variable
list of the JFNK solver. With particle enslavement, memory requirements
of the nonlinear solver are dramatically reduced. Particle equations
of motion are orbit-averaged and evolve self-consistently with the
field. The kinetic-enslaved JFNK not only makes the fully implicit
PIC algorithm practical, but also makes the fluid preconditioning
of the algorithm possible. 

It is worth pointing out that the preconditioned JFNK approach proposed
here can be conceptually viewed as an optimal combination of the direct
implicit and moment implicit approaches. The fluid preconditioner
is derived by taking the first two moments of the Vlasov equation,
and then linearizing them into a so-called ``delta-form'' \citep{knoll2004jacobian}.
Textbook linear analysis shows that such a system includes stiff electron
modes in an electrostatic plasma. Although taking large timesteps
for low-frequency field evolutions is desirable, previous work \citep{chen-jcp-11-ipic}
indicates that the implicit CPU speedup over explicit PIC is largely
insensitive to the timestep size for large enough timesteps owing
to particle sub-cycling for orbit resolution. It is thus sufficient
in this context to target the stiffest time scales supported, i.e.,
electron time scales. Therefore, we base our fluid preconditioner
on electron moment equations only. The implicit timestep is chosen
to resolve the ion plasma wave frequency. This is to resolve ion waves
of all scales (including the Debye length scale which is physically
relevant for some non-linear ion waves). For consistency with the
orbit averaging of the kinetic solver, we take the time-average of
the linearized moment equations in the preconditioner. We show that
the fluid preconditioner is asymptotic preserving in the sense that
it is well behaved in the quasineutral limit (as in Ref. \citep{degond-jcp-10-ap_pic}).
However, beyond the study in Ref. \citep{degond-jcp-10-ap_pic}, the
algorithm proposed here is also well behaved for arbitrary electron-ion
mass ratios. 

The rest of the paper is organized as follows. Section \ref{sec:Kinetic-enslavement}
motivates and introduces the concept to kinetic enslavement in the
implicit PIC formulation. Section \ref{sec:JFNK-method} introduces
the mechanics of the JFNK method and preconditioning. Section \ref{sec:formulation-pc}
formulates the fluid preconditioner of an electrostatic plasma system
in detail, with an extension to 1D non-uniform meshes. Linear analysis
of electron and ion waves, together with an asymptotic analysis of
the preconditioner are also provided. Section \ref{sec:Numerical-experiments}
presents numerical parametric experiments to test the performance
of the preconditioner. Finally, we conclude in Sec. \ref{sec:Conclusion}.

\section{Kinetically Enslaved Implicit PIC \label{sec:Kinetic-enslavement}}

We consider a collisionless electrostatic plasma system (without magnetic
field) described by the Vlasov-Ampere equations in one dimension (1D)
in both position ($x$) and velocity ($v$) \citep{chen-jcp-11-ipic}:
\begin{eqnarray}
\frac{\partial f_{\alpha}}{\partial t}+v\frac{\partial f_{\alpha}}{\partial x}+\frac{q_{\alpha}}{m_{\alpha}}E\frac{\partial f_{\alpha}}{\partial v} & = & 0,\label{eq:vlasov}\\
\epsilon_{0}\frac{\partial E}{\partial t}+j & = & \left\langle j\right\rangle ,\label{eq:ampere}
\end{eqnarray}
where $f_{\alpha}(x,v)$ is the particle distribution function of
species $\alpha$ in phase space, $q_{\alpha}$ and $m_{\alpha}$
are the species charge and mass respectively, $E$ is the self-consistent
electric field, $j$ is the current density, $\left\langle j\right\rangle =\int jdx/\int dx$,
and $\epsilon_{0}$ is the vacuum permittivity. The evolution of Vlasov
equation is solved by the method of characteristics, represented by
particles evolving according to Newton's equations of motion, 
\begin{eqnarray}
\frac{dx_{p}}{dt} & = & v_{p},\label{eq:dxpdt}\\
\frac{dv_{p}}{dt} & = & a_{p}.\label{eq:dvpdt}
\end{eqnarray}
Here $x_{p}$, $v_{p}$, $a_{p}$ are the particle position, velocity,
and acceleration, respectively, and $t$ denotes time. As a starting
point, we may discretize Eqs. \ref{eq:ampere}, \ref{eq:dxpdt}, and
\ref{eq:dvpdt} by a time-centered finite-difference scheme, to find:
\begin{eqnarray}
\epsilon_{0}\frac{E_{i}^{n+1}-E_{i}^{n}}{\Delta t}+j_{i}^{n+1/2} & = & \left\langle j\right\rangle ,\label{eq:discete-Ampere}\\
\frac{x_{p}^{n+1}-x_{p}^{n}}{\Delta t}-v_{p}^{n+1/2} & = & 0,\label{eq:discrete-xp}\\
\frac{v_{p}^{n+1}-v_{p}^{n}}{\Delta t}-\frac{q_{p}}{m_{p}}E(x_{p}^{n+1/2}) & = & 0,\label{eq:discrete-vp}
\end{eqnarray}
where a variable at time level $n+\nicefrac{1}{2}$ is obtained by
the arithmetic mean of the variable at $n$ and $n+1$, the subscript
$i$ denotes grid-index and subscript $p$ denotes particle-index,
$j_{i}=\sum_{p}q_{p}v_{p}S(x_{p}-x_{i})$, $E(x_{p})=\sum_{i}E_{i}S(x_{i}-x_{p})$,
and $S$ is a B-spline shape function \citep{christensen2010functions}. 

It is critical to realize that solving the complete system of field-particle
equations (i.e., with the field and particle position and velocity
as unknowns) in a Newton-Krylov-based solver is impractical, due to
the excessive memory requirements of building the required Krylov
subspace. To overcome the memory challenges of JFNK for implicit PIC,
the concept of kinetic enslavement has been introduced \citep{chen-jcp-11-ipic,markidis2011energy}.
With kinetic enslavement, the JFNK residual is formulated in terms
of the field equation only, nonlinearly eliminating Eqs. \ref{eq:discrete-xp}
and \ref{eq:discrete-vp} as auxiliary computations. The resulting
JFNK implementation has memory requirements comparable to that of
a fluid calculation. A single copy of particle quantities is still
needed for the required particle computations. 

One important implication of kinetic enslavement is that the enslaved
particle pusher has the freedom of being adaptive in its implementation.
This can be effectively exploited to overcome the accuracy shortcomings
of using a fixed timestep $\Delta t$ to discretize the time derivatives
of both field and particle equations \citep{Parker-jcp-93-bounded-multiscale-pic}.
This is so because solving low-frequency field equations demands using
large timesteps, but if particle orbits are computed with such timesteps,
large plasma response errors result \citep{langdon-jcp-79-pic_ts}.
In Ref. \citep{chen-jcp-11-ipic}, a self-adaptive, charge-and-energy-conserving
particle mover was developed that provided simultaneously accuracy
and efficiency. F\textcolor{black}{or each field timestep $\Delta t$,
the orbit integration step consists of four main }algorithmic elements:
\begin{enumerate}
\item Estimate the sub-timestep $\Delta\tau$ using a second order estimator
\citep{chen2013analytical}. 
\item Integrate the orbit over $\Delta\tau$ using a Crank-Nicolson scheme.
\item If a particle orbit crosses a cell boundary, make it land at the first
encountered boundary.
\item Accumulate the particle moments to the grid points.
\end{enumerate}
In the last step, the current density is orbit-averaged (over $\Delta t=\sum\Delta\tau$)
to ensure global energy conservation. Additionally, binomial smoothing
can be introduced without breaking energy or charge conservation.
This is done in the particle pusher by using the binomially smoothed
electric field, and the binomially smoothed orbit averaged current
density in Ampere's equation. The resulting Ampere's equation reads:
\begin{equation}
\epsilon_{0}\frac{E_{i}^{n+1}-E_{i}^{n}}{\Delta t}+SM(\bar{j})_{i}^{n+1/2}=\left\langle \overline{j}\right\rangle ^{n+1/2},\label{eq:bi-amperelaw}
\end{equation}
where the orbit averaged current density is: 
\begin{equation}
\overline{j}_{i}^{n+1/2}=\frac{1}{\Delta t\Delta x}\sum_{p}\sum_{\nu=1}^{N_{\nu}}q_{p}S(x_{i}-x_{p}^{\nu+1/2})v_{p}^{\nu+1/2}\Delta\tau^{\nu}.\label{eq:javerage}
\end{equation}
The binomial operator $SM$ is defined as $SM(Q)_{i}=\frac{Q_{i-1}+2Q_{i}+Q_{i+1}}{4}.$
A detailed description of the algorithm can be found in Ref. \citep{chen-jcp-11-ipic,chen-jcp-12-ipic_gpu}.

The kinetically enslaved JFNK residual is defined from Eq. \ref{eq:bi-amperelaw}
as: 
\begin{equation}
G_{i}(E^{n+1})=E_{i}^{n+1}-E_{i}^{n}+\frac{\Delta t}{\epsilon_{0}}\left(SM(\bar{j}[E^{n+1}])_{i}^{n+1/2}-\left\langle \overline{j}\right\rangle ^{n+1/2}\right).\label{eq:bi-amperelaw-residual}
\end{equation}
The functional dependence of $\bar{j}$ with respect to $E^{n+1}$
has been made explicit. Evaluation of $\bar{j}[E^{n+1}]$ requires
one particle integration step, and each linear and nonlinear iteration
of the JFNK method requires one residual evaluation . We summarize
the main elements of the JFNK nonlinear solver next.

\section{The JFNK solver\label{sec:JFNK-method}}

In its outer loop, JFNK employs Newton-Raphson's method to solve a
nonlinear system $\mathbf{G}(\mathbf{x})=0$, where $\mathbf{x}$
is the unknown, by linearizing the residual and inverting linear systems
of the form:
\begin{equation}
\left.\frac{\partial\mathbf{G}}{\partial\mathbf{x}}\right|^{(k)}\delta\mathbf{x}^{(k)}=-\mathbf{G}(\mathbf{x}^{(k)}),\label{eq:Newton-Raphson step}
\end{equation}
with $\mathbf{x}^{(k+1)}=\mathbf{x}^{(k)}+\delta\mathbf{x}^{(k)}$,
and $(k)$ denotes the nonlinear iteration number. Nonlinear convergence
is reached when:
\begin{equation}
\left\Vert \mathbf{G}(\mathbf{x}^{(k)})\right\Vert _{2}<\epsilon_{t}=\epsilon_{a}+\epsilon_{r}\left\Vert \mathbf{G}(\mathbf{x}^{(0)})\right\Vert _{2},\label{eq-Newton-conv-tol}
\end{equation}
where $\left\Vert \cdot\right\Vert _{2}$ is the Euclidean norm, $\epsilon_{t}$
is the total tolerance, $\epsilon_{a}$ is an absolute tolerance,
$\epsilon_{r}$ is the Newton relative convergence tolerance, and
$\mathbf{G}(\mathbf{x}^{(0)})$ is the initial residual.

Such linear systems are solved iteratively with a Krylov subspace
method (e.g. GMRES), which only requires matrix-vector products to
proceed. Because the linear system matrix is a Jacobian matrix, matrix-vector
products can be implemented Jacobian-free using the Gateaux derivative:
\begin{equation}
\left.\frac{\partial\mathbf{G}}{\partial\mathbf{x}}\right|^{(k)}\mathbf{v}=\lim_{\epsilon\rightarrow0}\frac{\mathbf{G}(\mathbf{x}^{(k)}+\epsilon\mathbf{v})-\mathbf{G}(\mathbf{x}^{(k)})}{\epsilon},\label{eq:gateaux}
\end{equation}
where $\mathbf{v}$ is a Krylov vector, and $\epsilon$ is in practice
a small but finite number (p. 79 in \citep{kelley1987iterative}).
Thus, the evaluation of the Jacobian-vector product only requires
the function evaluation $\mathbf{G}(\mathbf{x}^{(k)}+\epsilon\mathbf{v})$,
and there is no need to form or store the Jacobian matrix. This, in
turn, allows for a memory-efficient implementation. 

An inexact Newton method \citep{inexact-newton} is used to adjust
the convergence tolerance of the Krylov method at every Newton iteration
according to the size of the current Newton residual, as follows:
\begin{equation}
\left\Vert J^{(k)}\delta\mathbf{x}^{(k)}+\mathbf{G}(\mathbf{x}^{(k)})\right\Vert _{2}<\zeta^{(k)}\left\Vert \mathbf{G}(\mathbf{x}^{(k)})\right\Vert _{2}\label{eq-inexact-newton}
\end{equation}
where $\zeta^{(k)}$ is the inexact Newton parameter and $J^{(k)}=\left.\frac{\partial\mathbf{G}}{\partial\mathbf{x}}\right|^{(k)}$
is the Jacobian matrix. Thus, the convergence tolerance of the Krylov
method is loose when the Newton state vector $\mathbf{x}^{(k)}$ is
far from the nonlinear solution, and tightens as $\mathbf{x}^{(k)}$
approaches the solution. Superlinear convergence rates of the inexact
Newton method are possible if the sequence of $\zeta^{(k)}$ is chosen
properly (p. 105 in \citep{kelley1987iterative}). Here, we employ
the prescription:
\begin{eqnarray*}
\zeta^{A(k)} & = & \gamma\left(\frac{\left\Vert \mathbf{G}(\mathbf{x}^{(k)})\right\Vert _{2}}{\left\Vert \mathbf{G}(\mathbf{x}^{(k-1)})\right\Vert _{2}}\right)^{\alpha},\\
\zeta^{B(k)} & = & \min[\zeta_{max},\max(\zeta^{A(k)},\gamma\zeta^{\alpha(k-1)})],\\
\zeta^{(k)} & = & \min[\zeta_{max},\max(\zeta^{B(k)},\gamma\frac{\epsilon_{t}}{\left\Vert \mathbf{G}(\mathbf{x}^{(k)})\right\Vert _{2}})],
\end{eqnarray*}
with $\alpha=1.5$ , $\gamma=0.9$, and $\zeta_{max}=0.2$. The convergence
tolerance $\epsilon_{t}$ is defined in Eq. \ref{eq-Newton-conv-tol}.
In this prescription, the first step ensures superlinear convergence
(for $\alpha>1$), the second avoids volatile decreases in $\zeta_{k}$,
and the last avoids oversolving in the last Newton iteration. 

The Jacobian system Eq. \ref{eq:Newton-Raphson step} must be preconditioned
for efficiency. Here, we employ right preconditioning, which transforms
the original system into the equivalent one: 
\begin{equation}
JP^{-1}\mathbf{y}=-\mathbf{G}(\mathbf{x})
\end{equation}
where $J=\partial\mathbf{G}/\partial\mathbf{x}$ is the Jacobian matrix,
$P$ is a preconditioner, and $\delta\mathbf{x}=P^{-1}\mathbf{y}$.
The Jacobian-free preconditioned system employs
\begin{equation}
\mathit{J}P^{-1}\mathbf{v}=\lim_{\epsilon\rightarrow0}\frac{\mathbf{G}(\mathbf{x}+\epsilon P^{-1}\mathbf{v})-\mathbf{G}(\mathbf{x})}{\epsilon}\label{eq:gateaux-prec}
\end{equation}
for each Jacobian-vector product. An important feature of preconditioning
is that, while it may substantially improve the convergence properties
of the Krylov iteration (when $P$ approximates $J$ and is relatively
easy to invert), it does not alter the solution of the system upon
convergence. 

The purpose of this study is to formulate an effective, fast preconditioner
$P$ for the implicit PIC kinetic system. Before deriving the preconditioner,
however, we review the fundamental CPU speedup limits of implicit
vs. explicit PIC.

\section{Performance limits of implicit PIC}

As mentioned earlier, the ability of implicit PIC to take large timesteps
without numerical instabilities does not necessarily translate into
performance gains of implicit PIC over its explicit counterpart \citep{chen-jcp-11-ipic}.
In this section, we summarize the back-of-envelope estimate for the
CPU speedup introduced in the reference that supports this statement.

We begin by estimating the CPU cost for a given PIC solver to advance
the solution for a given time span $\Delta T$ as:
\begin{equation}
CPU=\frac{\Delta T}{\Delta t}N_{pc}\left(\frac{L}{\Delta x}\right)^{d}C,\label{eq:CPU-estimation-1}
\end{equation}
where $N_{pc}$ is the number of particles per cell, ($L/\Delta x$)
is the number of cells per dimension, $d$ is the number of physical
dimensions, and $C$ is the computational complexity of the solver
employed, measured in units of a standard explicit PIC Vlasov-Poisson
leap-frogd timestep. Accordingly, the implicit-to-explicit speedup
is given by:
\[
\frac{CPU_{ex}}{CPU_{im}}\sim\left(\frac{\Delta x_{im}}{\Delta x_{ex}}\right)^{d}\left(\frac{\Delta t}{\Delta t_{ex}}\right)\frac{1}{C_{im}},
\]
where we denote $\Delta t$ to be the implicit timestep. Assuming
that all particles take a fixed sub-timestep $\Delta\tau$ in the
implicit scheme, and that the cost of one timestep with the explicit
PIC solver is comparable to that of a single implicit sub-step, it
follows that $C_{im}\sim N_{FE}\left(\Delta t/\Delta\tau_{im}\right)$,
i.e., the cost of the implicit solver exceeds that of the explicit
solver by the number of function evaluations ($N_{FE}$) per $\Delta t$
multiplied by the number of particle sub-steps $\left(\Delta t/\Delta\tau_{im}\right)$.
Assuming typical values for $\Delta\tau_{im}\sim\min[0.1\Delta x/v_{th},\Delta t_{imp}]$,
$\Delta t_{ex}\sim0.1\omega_{pe}^{-1}$, $\Delta x_{im}\sim0.2/k$,
and $\Delta x_{ex}\sim\lambda_{D}$, we find that the CPU speedup
scales as:
\begin{equation}
\frac{CPU_{ex}}{CPU_{imp}}\sim\frac{0.2}{(5k\lambda_{D})^{d}}\min\left[\frac{1}{k\lambda_{D}},\sqrt{\frac{m_{i}}{m_{e}}}\right]\frac{1}{N_{FE}}.\label{eq:CPU-ex-im-1}
\end{equation}
This result supports two important conclusions. Firstly, it predicts
that the CPU speedup is asymptotically independent of the implicit
time step $\Delta t$ for $\Delta t\gg\Delta\tau_{im}$. The effect
of the implicit time step is captured in the extra power of one in
the $(k\lambda_{D})$ term, once one accounts for sub-stepping, but
that effect disappears when the mesh becomes coarse enough (i.e.,
$k\lambda_{D}<\sqrt{m_{e}/m_{i}}$). Also, it predicts that the speedup
improves with larger ion-to-electron mass ratio, indicating that the
approach is more advantageous when one employs realistic mass ratios.
Because the CPU speedup is asymptotically independent of $\Delta t$,
algorithmically it will be advantageous to use a time step that is
large enough to be in the asymptotic regime, but no larger. This will
motivate the choice in the preconditioner to include only electron
stiff physics.

Secondly, Eq. \ref{eq:CPU-ex-im-1} indicates that large CPU speedups
are possible when $k\lambda_{D}\ll1$, particularly in multiple dimensions,
but only if $N_{FE}$ is kept small and bounded. The latter point
motivates the development of suitable preconditioning strategies.
We focus on this in the next section.

\section{Fluid preconditioning the electrostatic implicit PIC kinetic system\label{sec:formulation-pc}}

The preconditioner of the nonlinear kinetic JFNK solver needs to return
an approximation for the $E$-field update only. The approximate $E$-field
update will be found from a linearized fluid model, consistently closed
with particle moments. As will be shown, the fluid model provides
an inexpensive approximation to the kinetic Jacobian. We demonstrate
the concept in the 1D electrostatic, multispecies PIC model.

\subsection{Formulation of the fluid preconditioner}

Following standard procedure \citep{knoll2004jacobian}, we work with
the linearized form of the governing equations to derive a suitable
preconditioner. The linearized, orbit-averaged, binomially smoothed
1D Ampere's residual equation (Eq. \ref{eq:discete-Ampere} with $E=E_{0}+\delta E$,
and $\delta\bar{j}\equiv\int_{0}^{\Delta t}\delta jdt/\Delta t$)
reads: 
\begin{equation}
\delta E=-\Delta t\left(G(E_{0})+\frac{1}{\varepsilon_{0}}SM(\delta\bar{j})\right),\label{eq:delta-Ampere-disc}
\end{equation}
where $G(E_{0})=E_{0}-E^{n}+\frac{\Delta t}{\varepsilon_{0}}(SM(\bar{j}_{0}^{n+\nicefrac{1}{2}})-\left\langle \bar{j}_{0}\right\rangle )$
is the residual of Ampere's law, the superscript $n$ denotes last
timestep, and the subscript 0 of the $E$-field denotes the current
Newton state. From the discussion in the previous section, for the
purpose of preconditioning we consider only the linear response of
electron contribution to the current ($\delta\bar{j}\simeq-e\delta\bar{\Gamma}$
where $\Gamma$ is the electron flux). Thus, the electric field update
in the preconditioner will be found from:
\begin{equation}
\delta E\approx-\Delta t\left(G(E_{0})-\frac{e}{\varepsilon_{0}}SM(\delta\bar{\Gamma})\right),\label{eq:delta-Ampere-e_only}
\end{equation}
where $\delta\bar{\Gamma}=\frac{1}{\Delta t}\int_{0}^{\Delta t}dt\delta\Gamma(t)$,
a time-average between timestep $n$ and $n+1$.. 

We approximate the linear response of the electron current via the
continuity and momentum equations of electrons, closed with moments
from particles (as in the implicit moment method \citep{mason-jcp-81-im_pic}).
The continuity equation for electrons is
\begin{equation}
\frac{\partial n}{\partial t}+\frac{\partial\Gamma}{\partial x}=0,\label{eq:continuity}
\end{equation}
where where $n$ is electron number density. Linearizing, we obtain:
\begin{equation}
\frac{\partial\delta n}{\partial t}=-\frac{\partial\delta\Gamma}{\partial x},\label{eq:delta-continuity}
\end{equation}
where we have used particle conservation ($\partial n_{0}/\partial t+\partial\Gamma_{0}/\partial x=0$),
which is satisfied at all iteration levels owing to exact charge conservation
\citep{chen-jcp-11-ipic}. We then take the time-average ($\frac{1}{\Delta t}\int_{0}^{\Delta t}dt$,
equivalently to the orbit average in Eq. \ref{eq:javerage}) of Eq.
\ref{eq:delta-continuity} to obtain 
\begin{equation}
\delta n=-\Delta t\frac{\partial\delta\bar{\Gamma}}{\partial x}.\label{eq:delta-continuity-disc}
\end{equation}
 The update equation for $\delta\bar{\Gamma}$ is found from the electron
momentum equation, which in conservative form reads 
\begin{equation}
m\left[\frac{\partial\Gamma}{\partial t}+\frac{\partial}{\partial x}\left(\frac{\Gamma\Gamma}{n}\right)\right]=-enE-\frac{\partial P}{\partial x}\label{eq:momentum}
\end{equation}
where $m$ is the electron mass, $P\equiv nT$ is the electron pressure,
and $T$ is the electron temperature. Linearizing it, we obtain:
\begin{equation}
m\left[\frac{\partial\delta\Gamma}{\partial t}+\frac{\partial}{\partial x}\left(\frac{2\Gamma_{0}\delta\Gamma}{n_{0}}-\frac{\Gamma_{0}\Gamma_{0}}{n_{0}^{2}}\delta n\right)\right]+e(n_{0}\delta E+\delta nE_{0})+\frac{\partial(\delta nT_{0})}{\partial x}=0,\label{eq:delta-momentum}
\end{equation}
where $T_{0}\equiv\int f(v)m(v-u)(v-u)dv/n_{0}$ is the current temperature
(or normalized pressure). Closures for $\Gamma_{0}$, $n_{0}$ and
$T_{0}$ are obtained from current particle information. In Eq. \ref{eq:delta-momentum},
we take $m\left[\partial\Gamma_{0}/\partial t+\partial(\Gamma_{0}\Gamma_{0}/n_{0})/\partial x\right]+en_{0}E_{0}+\partial(n_{0}T_{0})/\partial x=0$
by ansatz. To close the fluid model, we have neglected the linear
temperature response $\delta T$.

To cast Eq. \ref{eq:delta-momentum} in a useful form, we take its
time-derivative to get (assuming that $n_{0}$, $E_{0}$, and $T_{0}$
do not vary with time): 
\begin{equation}
m\frac{\partial^{2}\delta\Gamma}{\partial t^{2}}+e(n_{0}\frac{\partial\delta E}{\partial t}+\frac{\partial\delta n}{\partial t}E_{0})+\frac{\partial}{\partial x}(T_{0}\frac{\partial\delta n}{\partial t})=0,\label{eq:delta-momentum-dt}
\end{equation}
and then time-average the result to find (substituting Eqs. \ref{eq:delta-Ampere-disc}
and \ref{eq:delta-continuity}): 
\begin{equation}
\frac{2m\delta\bar{\Gamma}}{\Delta t^{2}}+e^{2}n_{0}\delta\bar{\Gamma}-eE_{0}\frac{\partial\delta\bar{\Gamma}}{\partial x}-\frac{\partial}{\partial x}(T_{0}\frac{\partial\delta\bar{\Gamma}}{\partial x})=-n_{0}G(E_{0}).\label{eq:delta-momentum-avg}
\end{equation}
Here, we have neglected the convective term for simplicity, and approximated
the first time-derivative term as: 
\begin{equation}
\frac{\partial\delta\Gamma}{\partial t}\simeq\frac{2\delta\bar{\Gamma}}{\Delta t}\label{eq:ddGamma-dt}
\end{equation}
(which is exact if $\delta\Gamma(t)$ is linear with $t$). We discretize
Eq. \ref{eq:delta-momentum-avg} with space-centered finite differences,
resulting in a tridiagonal system, which we invert for $\delta\bar{\Gamma}$
using a direct solver. Finally, we substitute the solution of $\delta\bar{\Gamma}$
in Eq. \ref{eq:delta-Ampere-e_only} to find the $E$-field update.

\subsection{Extension to curvilinear meshes}

The fully implicit PIC algorithm has been recently extended to curvilinear
meshes \citep{chacon-jcp-13-curvpic}. In this section, we rewrite
the above fluid model on a 1D non-uniform mesh using a map $x=x(\xi)$.
In 1D, the curvilinear form of in Eqs. \ref{eq:continuity} and \ref{eq:momentum}
can be derived straightforwardly by replacing every $dx$ with $\mathcal{J}d\xi$,
where $\mathcal{J}\equiv dx/d\xi$ is the Jacobian. It follows that
the continuity equation in logical space is written as: 
\begin{equation}
\frac{\partial n}{\partial t}+\frac{1}{\mathcal{J}}\frac{\partial\Gamma}{\partial\xi}=0.\label{eq:continuity-curv}
\end{equation}
The transformed momentum equation is 
\begin{equation}
m\left[\frac{\partial\Gamma}{\partial t}+\frac{1}{\mathcal{J}}\frac{\partial}{\partial\xi}\left(\frac{\Gamma\Gamma}{n}\right)\right]=qnE-\frac{1}{\mathcal{J}}\frac{\partial P}{\partial\xi}.\label{eq:momentum-curv}
\end{equation}
Similar to the procedure described above, linearizing and discretizing
Eqs. \ref{eq:continuity-curv} and \ref{eq:momentum-curv} again results
in a tridiagonal system.

\subsection{Electrostatic wave dispersion relations}

\begin{figure}

\begin{centering}
\includegraphics[scale=0.7]{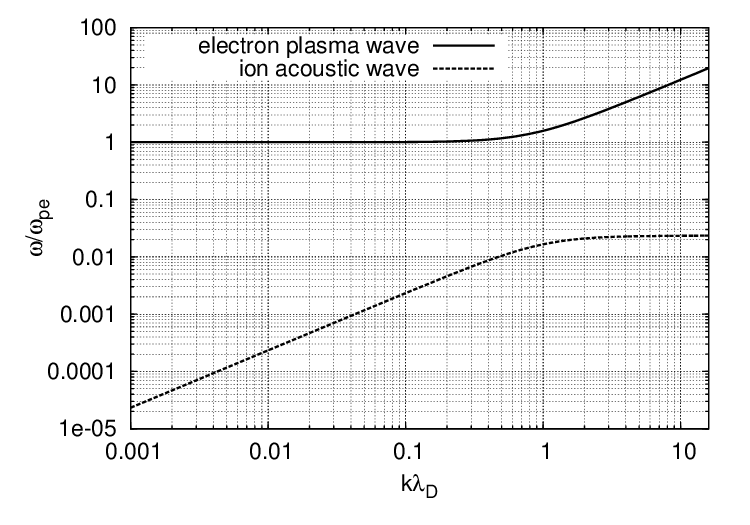}\caption{\label{fig:eiwave-disper}Dispersion relations of electron and ion
waves in an electrostatic plasma. The dispersions can be obtained
by Fourier analysis of the fluid model of Eq. \ref{eq:delta-Ampere-disc},
\ref{eq:delta-continuity} and \ref{eq:delta-momentum}, for both
electrons and ions, assuming that $E_{0},\Gamma_{0}$, $n_{0},T_{0}=$const.}

\par\end{centering}

\end{figure}

It is instructive to look at the dispersion relation of Eq. \ref{eq:delta-Ampere-disc},
\ref{eq:delta-continuity} and \ref{eq:delta-momentum}, for both
electrons and ions. Figure \ref{fig:eiwave-disper} shows the dispersion
relation of electron plasma waves and ion acoustic waves \citep{FChenbook},
from which we make the following observations. The stiffest wave is
the electron plasma wave, whose frequency $\omega_{pe}$ is essentially
insensitive to the wave number $k$ for $k\lambda_{D}<1$. The wave
frequency increases for $k\lambda_{D}>1$, but in that range the plasma
wave is highly Landau-damped \citep{jackson1960longitudinal}. In
contrast to the electron wave, the ion wave frequency increases with
$k$ for $k\lambda_{D}<1$, but saturates at $\sim\omega_{pi}$ for
$k\lambda_{D}>1$. In a propagating ion acoustic wave (IAW), nonlinear
effects lead to wave steepening. Because of the wave dispersion, the
IAW steepening stops when the high frequency waves propagate slower
than the low frequency ones \citep{krall1997we}. Those high frequency
ion waves are physically important, and therefore need to be resolved.
For this reason, in our numerical experiments, we limit the implicit
time step to $\Delta t\sim0.1\omega_{pi}^{-1}$. The frequency gap
between the electron and ion waves is about a factor of $\sqrt{m_{i}/m_{e}}$,
which provides enough room to place the algorithm in the large timestep
asymptotic regime (Eq. \ref{eq:CPU-ex-im-1}).

\subsection{Asymptotic behavior of the implicit PIC formulation in the quasineutral
limit.\label{sub:Asymptotic-preservation-in}}

Since the implicit scheme is able to use large grid sizes and timesteps
stably, it is important to ensure that the fluid preconditioner be
able to capture relevant asymptotic regimes correctly \citep{degond-jcp-10-ap_pic}.
In the context of electrostatic PIC, the relevant asymptotic regime
is the quasineutral limit, which manifests when the domain length
is much larger than the Debye length ($L\gg\lambda_{D}$) and when
$m_{e}\ll m_{i}$. In this limit, the electric field must be found
from the fluid equations \citep{fernsler2005quasineutral}, and leads
to the well know ambipolar electric field, $E=-\frac{1}{en}\partial_{x}P$.

In our context, the algorithm must be well behaved when $L$ varies
from $\sim\lambda_{D}$ to $\gg\lambda_{D}$, and for arbitrary mass
ratios. In particular, the fluid preconditioner must feature these
properties to successfully accelerate the kinetic algorithm. To confirm
that this is the case, following Ref. \citep{degond-jcp-10-ap_pic}
we normalize the electron fluid equations to the following reference
quantities:
\begin{equation}
\hat{x}=\frac{x}{x_{0}},\:\hat{v}=\frac{v}{v_{0}},\:\hat{t}=\frac{tv_{0}}{x_{0}},\:\hat{n}=\frac{n}{n_{0}},\:\hat{q}=\frac{q}{q_{0}},\:\hat{m}=\frac{m}{m_{0}},\:\hat{E}=\frac{Eq_{0}x_{0}}{k_{B}T_{0}}.
\end{equation}
We choose $x_{0}=L$, $v_{0}=\sqrt{k_{B}T_{0}/m_{0}}$, $q_{0}=e$,
$m_{0}=m_{i}$. For electrons, $q=-e$, and hence $\hat{q}=-1$. The
normalized preconditioning equations become:
\begin{eqnarray}
\hat{\lambda}_{D}^{2}\frac{\partial\hat{E}}{\partial\hat{t}}-\hat{\Gamma} & = & 0,\label{eq:Ampere-law-norm}\\
\frac{\partial\hat{n}}{\partial\hat{t}}+\frac{\partial\hat{\Gamma}}{\partial\hat{x}} & = & 0,\label{eq:continuity-norm}\\
\hat{m}\frac{\partial\hat{\Gamma}}{\partial\hat{t}}+\hat{n}\hat{E}+\hat{T}\frac{\partial\hat{n}}{\partial\hat{x}} & = & 0,\label{eq:momentum-norm}
\end{eqnarray}
where in Eq. \ref{eq:momentum-norm} we have neglected the convective
term. Substituting Eq. \ref{eq:Ampere-law-norm} into Eq. \ref{eq:momentum-norm},
we find the equation for the electric field:
\begin{equation}
\hat{m}\frac{\partial}{\partial\hat{t}}\left(\hat{\lambda}_{D}^{2}\frac{\partial\hat{E}}{\partial\hat{t}}\right)+\hat{n}\hat{E}+\hat{T}\frac{\partial\hat{n}}{\partial\hat{x}}=0,\label{eq:ap-momentum}
\end{equation}
where $\lambda_{D}$ may change in time and space. The solution of
$\hat{E}$ is well behaved as $\hat{m}\hat{\lambda}_{D}^{2}\rightarrow0$,
where we indeed find that $\hat{n}\hat{E}=-\hat{T}\frac{\partial\hat{n}}{\partial\hat{x}}$,
which is the correct (ambipolar) $E$-field. Our fluid preconditioner
is based on the linearization of Eqs. \ref{eq:Ampere-law-norm}-\ref{eq:momentum-norm},
and therefore inherits this asymptotic property. In what follows,
we will demonstrate among other things the effectiveness of the preconditioner
as we vary the domain size and the mass ratio.

\section{Numerical experiments\label{sec:Numerical-experiments}}

We use the IAW problem for testing the performance of the fluid-based
preconditioner. IAW propagation is a multi-scale problem determined
by the coupling between electrons and ions. The 1D case used in Ref.~\citep{chen-jcp-11-ipic}
features large-amplitude IAWs in an unmagnetized, collisionless plasma
without significant damping. \textcolor{black}{The base simulation
parameters used here are $T_{e}/T_{i}=545$ and $m_{i}/m_{e}=1836$.
The periodic computational domain, measured in units of Debye length,
is discretized with both uniform and non-uniform meshes. We test the
solver performance by varing the timestep, the electron-ion-mass-ratio,
the domain length, the number of particles, and the number of cells,
with and without preconditioning. We also compare the solver performance
against explicit PIC simulations. In all cases, the nonlinear tolerances
of implicit JFNK solver are set to $\varepsilon_{r}=1\times10^{-8}$
and $\varepsilon_{a}=0$. The number of function evaluations (NFE),
given by the number of Newton-Raphson and GMRES iterations, is monitored
and averaged over 20 timesteps. With the proposed linear preconditioner,
the performance gain will stem mainly from reducing the GMRES iteration
count. The Newton iteration count remains nearly constant (at about
4 to 5 iterations, unless otherwise stated) regardless of preconditioning.}

We initialize the calculation with the following ion distribution
function: 
\begin{equation}
f(x,v,t=0)=f_{M}(v)\left[1+a\cos\left(\frac{2\pi}{L}x\right)\right]\label{eq:initf}
\end{equation}
where $f_{M}(v)$ is the Maxwellian distribution, $a$ is the perturbation
level, $L$ is the domain size. The spatial distribution is approximated
by first putting ions randomly with a constant distribution, e.g.
$x^{0}\in[0,L]$. The electrons are distributed in pairs with ions
according to the Debye distribution\citep{williamson1971initial}.
Specifically, in each $e$-$i$ pair, the electron is situated away
from the ion by a small distance, $dx=\mathrm{ln}(R)$ where $R\in(0,1)$
is a uniform random number (note that we normalize all lengths with
the electron Debye length). We then shift the particle position by
a small amount such that $x=x^{0}+a\cos\left(\frac{2\pi}{L}x^{0}\right)$,
with $a=0.2$.

For testing the solver performance with non-uniform meshes, the mesh
adaptation in the periodic domain is provided by the map \citep{chacon-jcp-13-curvpic}:
\begin{equation}
x(\xi)=\xi+\frac{L}{2\pi}(1-\frac{N\Delta x_{\nicefrac{L}{2}}}{L})\sin\left(\frac{2\pi\xi}{L}\right),\label{eq:map-xofxi}
\end{equation}
which has the property that the Jacobian $J$ is also periodic. Here,
$N$ is the number of mesh points, and $\Delta x_{\nicefrac{L}{2}}$
is the physical mesh resolution at $x=\xi=L/2$.

Before we begin the convergence studies, it is informative to look
at the condition number of the Jacobian system, which can be estimated
as the number of times we step over the explicit CFL: 
\begin{equation}
\sigma\propto\omega_{pe}\Delta t=0.1\frac{\omega_{pe}}{\omega_{pi}}=0.1\sqrt{\frac{m_{i}}{m_{e}}},\label{eq:condition_number}
\end{equation}
where we have used that $\Delta t\sim0.1\omega_{pi}^{-1}$, and we
have assumed $k\lambda_{D}<1$. The first important observation is
that, as expected, the Jacobian system will become harder to solve
as we increase the ion-to-electron mass ratio. Secondly, the condition
number does not depend on $k\lambda_{D}$. The latter, while surprising,
is a consequence of our chosen implicit time step upper bound. Dependence
of $\sigma$ with $k\lambda_{D}$ is recovered for $k\lambda_{D}>1$,
but in this regime Langmuir waves are highly Landau-damped \citep{jackson1960longitudinal},
and do not survive in the system.

\begin{figure}
\begin{centering}
\includegraphics{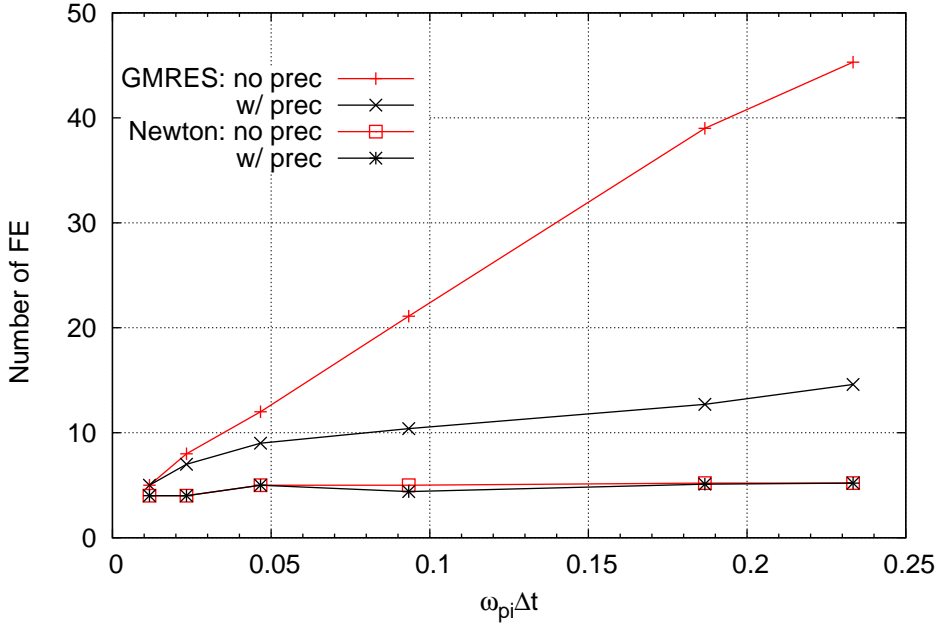}
\par\end{centering}

\caption{\label{fig:FE-dt}The performance of the JFNK solver against the timestep,
with $L=100$, $N_{x}=128$, $N_{pc}=1000$, and $m_{i}/m_{e}=1836$.
The number of function evaluations are well controlled by the preconditioner
over a large range of $\Delta t$. }
\end{figure}

We demonstrate the performance of the fluid preconditioner by varying
several relevant parameters, namely, the implicit timestep~$\Delta t$,
the mass ratio $m_{i}/m_{e}$, the domain length $L$, the mesh size
$N_{x}$, and the number of particles per cell $N_{pc}$. We begin
with the implicit timestep, which we vary from $0.01\omega_{pi}^{-1}$
to $0.25\omega_{pi}^{-1}$. For this test, we choose $L=100$, $N_{x}=128$,
$N_{pc}=1000$, and $m_{i}/m_{e}=1836$. As shown in Figure \ref{fig:FE-dt},
the performance for preconditioned and unpreconditioned solvers is
about the same for small time steps, where the Jacobian system is
not stiff. However, significant differences in performance develop
for larger timesteps, reaching a factor of 2 to 3 as the timestep
approaches $0.2\omega_{pi}^{-1}$. Overall, the preconditioner is
able to keep the linear and nonlinear iteration count fairly well
bounded as the timestep increases. 

\begin{figure}
\centering{}\includegraphics{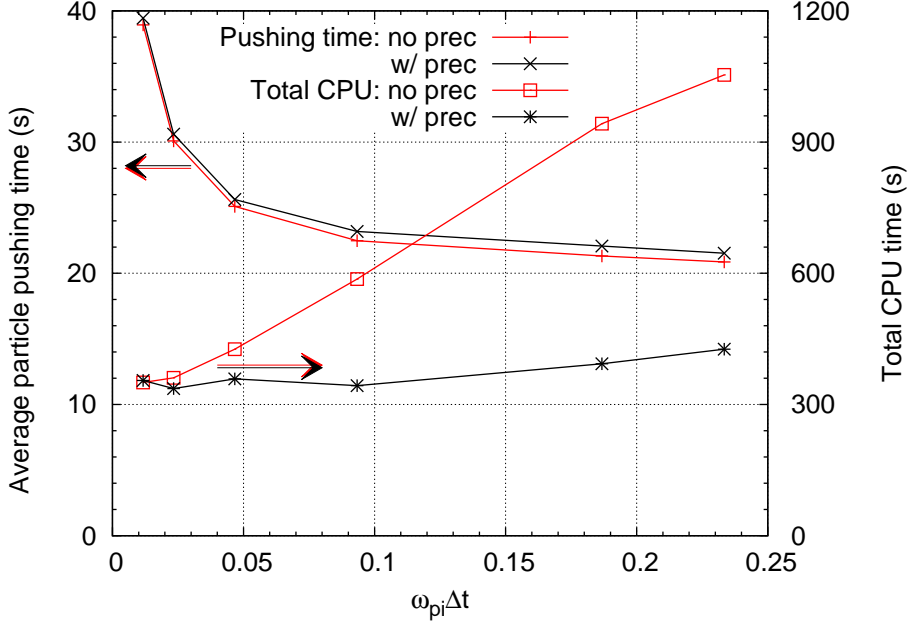}\caption{\label{fig:CPU-dt}Overall CPU performance as a function of timestep,
comparing the unpreconditioned and preconditioned solvers in terms
of the average particle pushing time (obtained by the total CPU time
divided by the average number of iterations) (left) and wall clock
CPU time (right). $L=100$, $N_{x}=128$, $m_{i}/m_{e}=1836$, and
the time-span is fixed at 4.67$\omega_{pi}^{-1}$ for all computations.}
\end{figure}

For bounded $N_{FE}$, Eq. \ref{eq:CPU-ex-im-1} predicts that the
actual CPU time should be largely insensitive to the timestep size.
This is confirmed in Fig. \ref{fig:CPU-dt}, which shows the CPU performance
of a series of computations with a fixed simulation time-span. Clearly,
the total CPU time is essentially independent of $\Delta t$ with
preconditioning (but not without). Also, both with and without preconditioning,
the average particle pushing time, which is the average CPU time used
for all particle pushes during the simulation time-span, saturates
for large enough time steps (e.g. $v_{the}\Delta t>1\sim10\Delta x$),
indicating that we have reached an asymptotically large time step.
Even though the CPU performance of the preconditioned solver is independent
of $\Delta t$, the use of larger timesteps is beneficial for the
following reasons. Firstly, the orbit-averaging performed to obtain
the plasma current density helps with noise reduction, as it provides
the time average of many samplings per particle \citep{cohen-jcp-82-orbit_averaging}.
Secondly, the operational intensity (computations per memory operation)
per particle orbit increases with the timestep, which helps enhance
the computing performance (or efficiency) and offset communication
latencies in the simulation \citep{chen-jcp-12-ipic_gpu}.

The performance of the preconditioner vs. the electron-ion mass ratio
for both uniform and non-uniform meshes is shown in Tables \ref{tab:ES-prec-performance-iaw-uniform}
and \ref{tab:ES-prec-performance-IAW-non-uniform}. To make a fair
comparison, both uniform and non-uniform meshes have the same finest
mesh resolution, which locally resolves the Debye length. From the
tables it is clear that similar performance gains of the preconditioned
solver vs. the unpreconditioned one are obtained for both uniform
and non-uniform meshes. The dependence of the GMRES performance on
the mass ratio is much weaker with the preconditioner: as the mass
ratio increases by a factor of 100, the GMRES iteration count increases
by a factor of 5 without the preconditioner, vs. a factor of 2 with
the preconditioner. Although not completely independent of the mass
ratio, the solver behavior is consistent with the asymptotic analysis
made in Sec. \ref{sub:Asymptotic-preservation-in}.

\begin{table}
\centering{}\caption{\label{tab:ES-prec-performance-iaw-uniform}Solver performance with
and without the fluid preconditioner for the IAW case with $L=100$,
$N_{x}=512$, and $N_{pc}=1000$ on a uniform mesh. For all the test
cases, $\Delta t=0.1\omega_{pi}^{-1}$. The Newton and GMRES iteration
numbers are obtained by an average over 20 timesteps. For all the
runs, we have kept the ion and electron temperature constant.}
\begin{tabular}{|c|c|c|c|c|}
\hline 
\multirow{2}{*}{$m_{i}/m_{e}$} & \multicolumn{2}{c|}{no preconditioner} & \multicolumn{2}{c|}{with preconditioner}\tabularnewline
\cline{2-5} 
 & Newton  & GMRES & Newton  & GMRES\tabularnewline
\hline 
100 & 4 & 8 & 4 & 7\tabularnewline
\hline 
1600 & 5 & 21.2 & 4 & 10.1\tabularnewline
\hline 
10000 & 5.8 & 50.1 & 5.5 & 13.5\tabularnewline
\hline 
\end{tabular}
\end{table}

\begin{table}
\centering{}\caption{\label{tab:ES-prec-performance-IAW-non-uniform}Solver performance
with and without the fluid preconditioner for the IAW case with the
non-uniform mesh ($N_{x}=64$ and the smallest mesh size 0.2).}
\begin{tabular}{|c|c|c|c|c|}
\hline 
\multirow{2}{*}{$m_{i}/m_{e}$} & \multicolumn{2}{c|}{no preconditioner} & \multicolumn{2}{c|}{with preconditioner}\tabularnewline
\cline{2-5} 
 & Newton  & GMRES & Newton  & GMRES\tabularnewline
\hline 
100 & 4 & 7.6 & 4 & 7\tabularnewline
\hline 
1600 & 5 & 21.3 & 5.1 & 12.1\tabularnewline
\hline 
10000 & 5.8 & 48.6 & 5.3 & 16.5\tabularnewline
\hline 
\end{tabular}
\end{table}

\begin{figure}
\centering{}\includegraphics{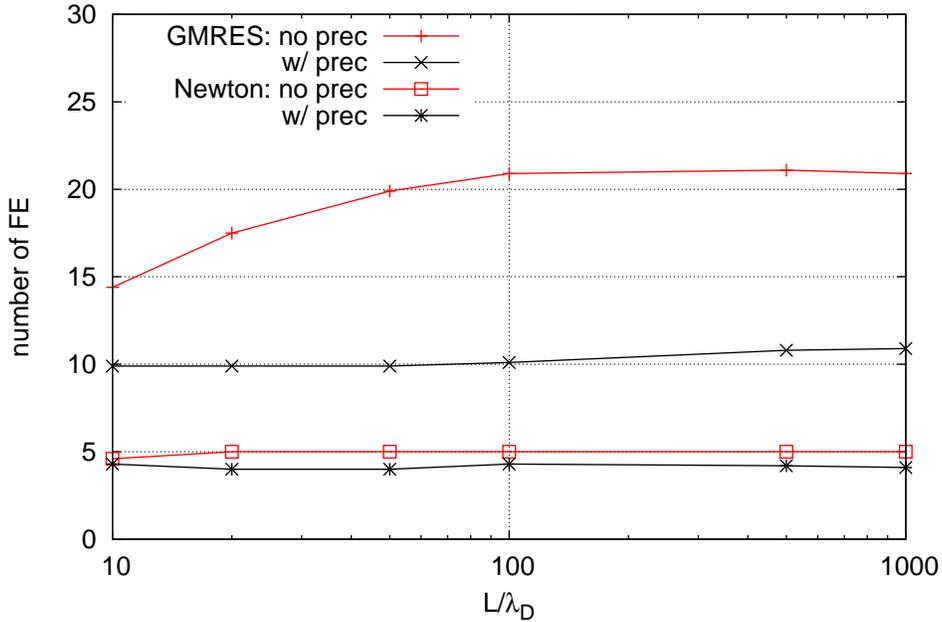}\caption{\label{fig:FE_L_iaw}Solver performance as a function of the domain
size, with $N_{x}=64$, $N_{pc}=1000$, $\Delta t=0.1\omega_{pi}^{-1}$.}
\end{figure}

The impact of the domain length in the solver performance is shown
in Fig. \ref{fig:FE_L_iaw}. Clearly, the solver performance remains
fairly insensitive to the domain length both with and without the
preconditioner, even though the domain length varies from 10 to 1000
Debye lengths. This is consistent with the condition number analysis
in Eq. \ref{eq:condition_number}. The impact of the preconditioner
in the number of GMRES iterations is expected for the time step chosen.

\begin{figure}
\begin{centering}
\includegraphics{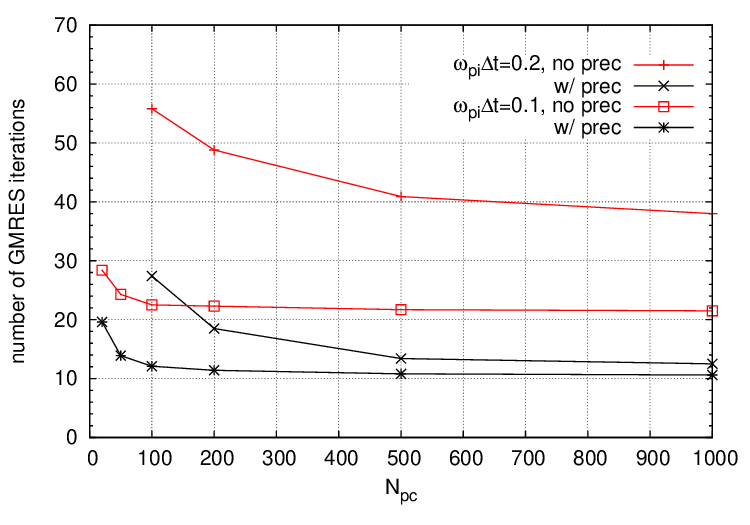}
\par\end{centering}

\begin{centering}
\includegraphics{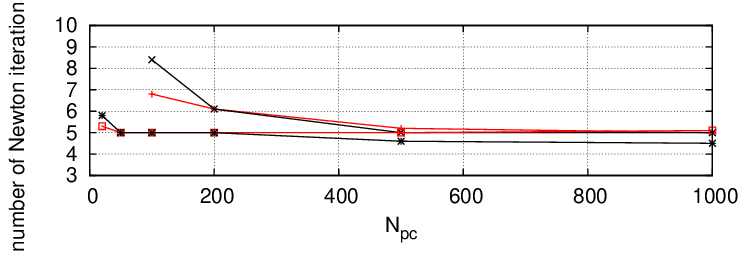}
\par\end{centering}

\caption{\label{fig:FE-particle}NFE of GMRES and Newton iterations as a function
of average number of particles per cell for a domain size $L=100$
with $N_{x}=128$ uniformly distributed cells. }
\end{figure}

The impact of the number of particles in the performance of the solver
is shown in Fig. \ref{fig:FE-particle}, which depicts the iteration
count of both Newton and GMRES vs. the number of particles. The timestep
is varied by a factor of two, corresponding to about one-tenth and
one-fifth of the inverse ion plasma frequency ($\omega_{pi}^{-1}$).
As expected, the solver performs better with smaller timesteps and
with larger number of particles. The number of linear and nonlinear
iterations increases as the number of particles decreases. This behavior
is likely caused by the increased interpolation noise associated with
fewer particles: the noise in charge density results in fluctuations
in the self-consistent electric field, making the Jacobian-related
calculations less accurate, thus delaying convergence. The preconditioner
seems to ameliorate the impact of having too few particles on the
performance of the algorithm, thus robustifying the nonlinear solver. 

\begin{figure}
\centering{}\includegraphics{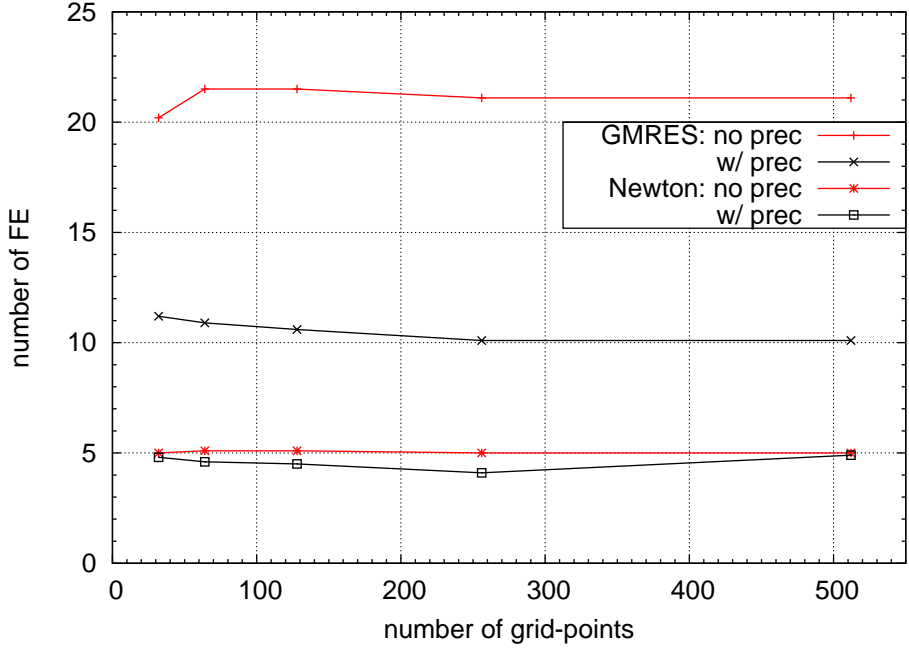}\caption{\label{fig:FE-dx}Solver performance vs. the number of grid-points
$N_{x}$ for $\omega_{pi}\Delta t=0.093$, $L=100\lambda_{D}$, and
$N_{pc}=1000$.}
\end{figure}

\begin{figure}
\centering{}\includegraphics{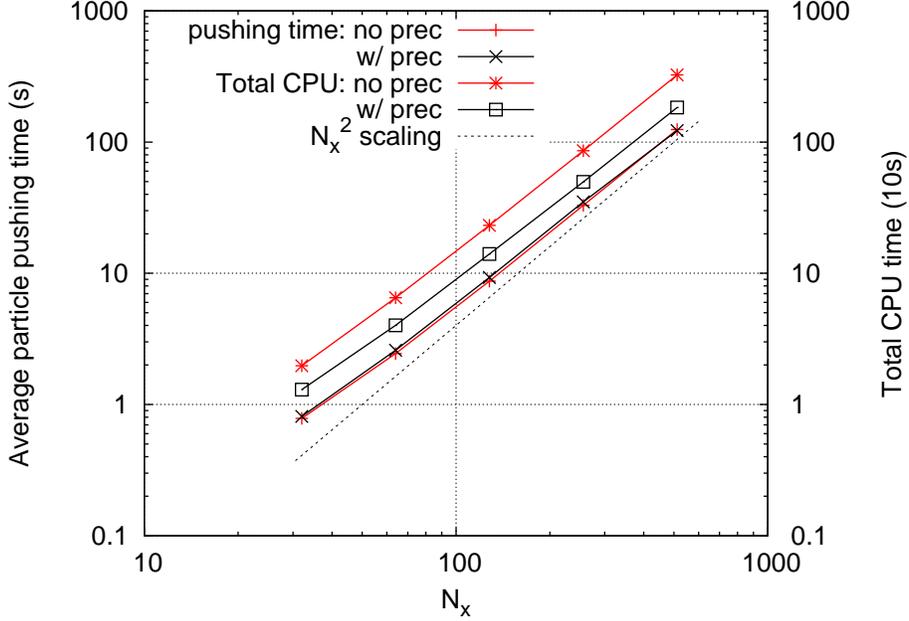}\caption{\label{fig:CPU-nx}The performance of the JFNK solver against the
number of grid-points, with $N_{pc}=1000$. The average particle pushing
time is shown on the left and total CPU time for a total time-span
$80$ is shown on the right. Both particle pushing time and total
CPU time scale as $N_{x}^{2}$ for large enough $N_{x}$.}
\end{figure}

The impact of the number of grid points on the solver performance
is shown in Fig. \ref{fig:FE-dx} for $\omega_{pi}\Delta t=0.093$,
$L=100\lambda_{D}$, and $N_{pc}=1000$. We see that the linear and
nonlinear iteration count remains fairly constant with respect to
$N_{x}$, with and without preconditioning. This is consistent with
the condition number result in Eq. \ref{eq:condition_number} (which
is independent of the wavenumber). However, despite the fact that
the number of iterations is virtually independent of the number of
grid points, the CPU time grows significantly with it. Figure \ref{fig:CPU-nx}
shows that the computational cost scales as $N_{x}^{2}$ for $N_{x}$
large enough. The reason is two-fold. On the one hand, since we keep
the number of particles per cell fixed, the computational cost of
pushing particles increases proportionally with the number of grid
points. On the other hand, as we refine the grid, the cost per particle
increases because particles have to cross more cells (for a given
timestep). In multiple dimensions, the particle orbit will sample
$N^{1/d}$ cells on average, for large enough $N$ (or $\Delta t$),
with $N$ and $d$ denoting the total number of grid points and dimensions,
respectively. Hence, the cost of particle crossing will scale as $N^{1/d}$,
and the computational cost will scale as $N^{1+1/d}$. In this sense,
the 1D configuration is the least favorable.

\begin{figure}
\begin{centering}
\includegraphics{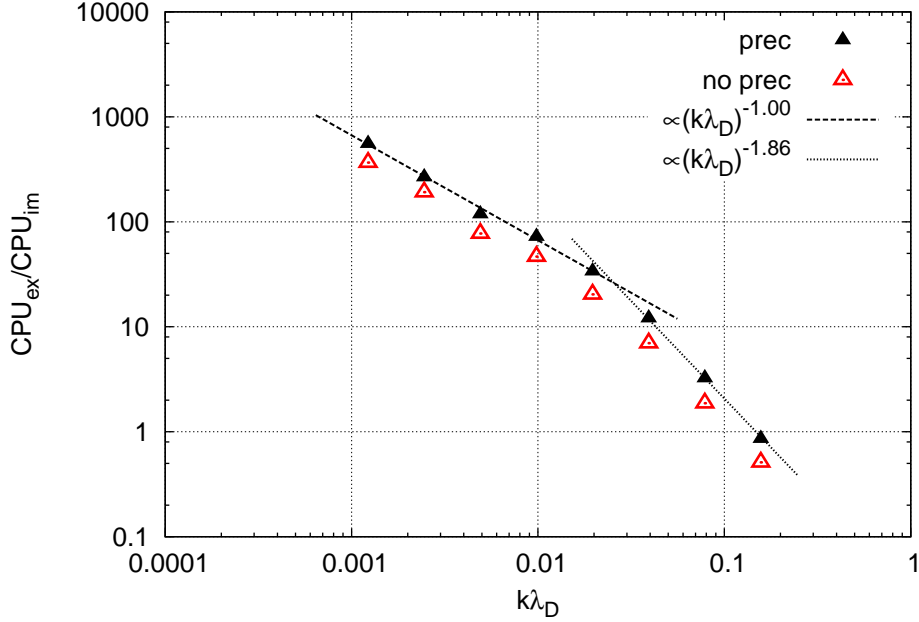}
\par\end{centering}

\caption{\label{fig:CPU-scaling}The implicit PIC solver performance compared
with the explicit scheme. The performance gain increases with the
domain size. For the parameters used, the performance gain of the
implicit solver is enhanced by the preconditioner by about a factor
of 2.}
\end{figure}

The performance of the implicit PIC solver vs. the explicit PIC one
is compared in Fig. \ref{fig:CPU-scaling}, which depicts the CPU
speedup vs. $k\lambda_{D}$. For this test, we choose $m_{i}/m_{e}=1836$,
$\Delta t=0.1\omega_{pi}^{-1}$, and $N_{pc}=1000$. In the implicit
tests, the number of grid-points is kept fixed at $N_{x}=32$ as $L$
increases with $k=2\pi/L$. In the explicit computations, $\Delta x\simeq0.3\lambda_{D}$
is kept constant for stability, and therefore the number of grid-points
increases with $L$. Both implicit and explicit tests employ a uniform
mesh. We monitor the scaling power index of Eq. \ref{eq:CPU-ex-im-1}
with and without the preconditioner. We test the performance with
a large implicit timestep (about 40 times larger than the explicit
timestep). The scaling index is found to be $\sim1.86$ for small
domain sizes, close to the expected value of 2. As $L$ increases,
the scaling index becomes $\sim1$. The scaling index turns at $k\lambda_{D}\sim\sqrt{m_{e}/m_{i}}\sim0.025$,
as predicted by Eq. \ref{eq:CPU-ex-im-1}. The estimated scaling index
of 2 would be recovered if one increased the timestep proportionally
to $L$, but this would result in timesteps too large with respect
to $\omega_{pi}^{-1}$. Overall, these results are in very good agreement
with our simple estimates. The preconditioned solver gains about a
factor of two compared to the un-preconditioned one, insensitively
to $k\lambda_{D}$, which is consistent with the results depicted
in Fig. \ref{fig:FE_L_iaw}. We see that for $k\lambda_{D}<10^{-3}$,
the implicit scheme delivers speedups of about three orders of magnitude
vs. the explicit approach, while remaining exactly energy- and charge-conserving. 

The setup in Fig. \ref{fig:CPU-scaling} employs a uniform mesh. However,
sometimes it is necessary to resolve the Debye length locally, e.g.
at a shock front or a boundary layer near a wall. In this case, using
a non-uniform mesh is advantageous \citep{chacon-jcp-13-curvpic}.
We test the performance of the preconditioner on a non-uniform mesh
for a nonlinear ion acoustic shock wave, as setup in Ref. \citep{chacon-jcp-13-curvpic}.
Specifically, we use $L=100\lambda_{D}$, $N_{pc}=2000$, $N_{x}=64$,
and $\Delta t=0.1\omega_{pi}^{-1}$ for 20 timesteps. The minimum
resolution is $\Delta x=0.5\lambda_{D}$ at the shock location (as
in \citep{chacon-jcp-13-curvpic}, we perform the simulation in the
reference frame of the shock). With a nonlinear tolerance $\epsilon_{r}=2\times10^{-4}$,
we have found that, with preconditioning, the average number of Newton
and GMRES iterations is 3 and 10.1, respectively, compared to 3.6
and 23 without preconditioning. The performance gain in the linear
solve is about factor of two, comparable to that obtained for a uniform
mesh with similar problem parameters. Similar performance gains are
found with tighter nonlinear tolerances: for $\epsilon_{r}=10^{-8}$,
we find 5.1 Newton and 46.6 GMRES iterations without preconditioning,
vs. 5 and 20.5 with it. 

\section{Conclusions\label{sec:Conclusion}}

This study has focused on the development of a preconditioner for
a recently proposed fully implicit, JFNK-based, charge- and energy-conserving
particle-in-cell electrostatic kinetic model \citep{chen-jcp-11-ipic}.
In the reference, it was found that, for large enough implicit time
steps $\Delta t$, the potential implicit-to-explicit CPU speedup
scaled as $\frac{1}{N_{FE}(k\lambda_{D})^{d}}$, with $N_{FE}$ the
number of function evaluations per time step, and $k\lambda_{D}\propto\lambda_{D}/L$.
Thus, large speedups are expected when $k\lambda_{D}\ll1$ provided
that $N_{FE}$ is kept bounded. While the CPU speedup does not scale
directly with $\Delta t$, the use of large $\Delta t$ is advantageous
to maximize operational intensity \citep{chen-jcp-12-ipic_gpu} (i.e.,
to maximize floating point operations per byte communicated), and
to control numerical noise via orbit averaging \citep{cohen-jcp-82-orbit_averaging}.

We have targeted a preconditioner based on an electron fluid model,
which is sufficient to capture the stiffest time scales, and thus
enable the use of large implicit time steps while keeping the number
of function evaluations bounded. The performance of the preconditioned
kinetic JFNK solver has been analyzed with various parametric studies,
including time step, mass ratio, domain length, number of particles,
and mesh size. The number of function evaluations is found to be insensitive
against changes in all of these, delivering a robust nonlinear solver.
The CPU time of the implicit PIC solver is found to be insensitive
to the time step (as expected), but to scale with the square of the
number of mesh points in 1D. This scaling is due to the number of
particles per cell being kept constant, and to the number of particle
crossings increasing linearly with the mesh resolution. The latter
scaling will be more benign in multiple dimensions, as particle orbits
remain one-dimensional. Speedups of about three orders of magnitude
vs. explicit PIC are demonstrated when $\lambda_{D}\ll L$ (i.e.,
in the quasineutral regime). Based on the speedup prediction in \citep{chen-jcp-11-ipic},
more dramatic speedups are expected in multiple dimensions.

We conclude that the proposed algorithm shows much promise for extension
to multiple dimensions and to electromagnetic simulations. This will
be the subject of future work.

\paragraph*{Acknowledgments}

This work was partially sponsored by the Office of Fusion Energy Sciences
at Oak Ridge National Laboratory, and by the Los Alamos National Laboratory
(LANL) Directed Research and Development Program. This work was performed
under the auspices of the US Department of Energy at Oak Ridge National
Laboratory, managed by UT-Battelle, LLC under contract DE-AC05-00OR22725,
and the National Nuclear Security Administration of the U.S. Department
of Energy at Los Alamos National Laboratory, managed by LANS, LLC
under contract DE-AC52-06NA25396.

\pagebreak{}

\bibliographystyle{ieeetr}
\bibliography{kinetic}

\end{document}